\documentclass[aps,prl,twocolumn,sort&compress, floatfix,showpacs]{revtex4}
\usepackage{amsmath}
\usepackage{graphicx}

\def\Av#1{\langle#1\rangle}
\newcommand*{\auf}{\uparrow}
\newcommand*{\ab}{\downarrow}
\renewcommand*{\vec}[1]{\mathbf{#1}}

\begin{document}

\author{J. Unterhinninghofen} 
\email{Julia@fkf.mpg.de}
\author{D. Manske}%
\affiliation{Max-Planck Institute for Solid State
Research, Heisenbergstr. 1, 70569 Stuttgart, Germany}
\author{A. Knorr}
\affiliation{Technical University Berlin, Institute for Theoretical Physics, Hardenbergstr. 36, 10623 Berlin, Germany}
\date{\today} 

\title{Theory of ultrafast nonequilibrium dynamics in $d$-wave superconductors}

\begin{abstract}
  We use density-matrix theory to calculate the ultrafast dynamics of
  unconventional superconductors from a microscopic viewpoint. We calculate
  the time evolution of the optical conductivity as well as pump-probe spectra
  for a $d$-wave order parameter. Three regimes can be distinguished in the
  spectra. The Drude response at low photon energies is the only one of those
  which has been measured experimentally so far.  At higher energies, we
  predict two more regimes: the pair-breaking peak, which is reduced as
  Cooper-pairs are broken up by the exciting pulse; and a suppression above
  the pair-breaking peak due to nonequilibrium quasiparticles. Furthermore, we
  consider the influence of the electron-phonon coupling, and derive rate
  equations which have been widely used so far.
\end{abstract}

\maketitle

\textit{{\label{sec:introduction}Introduction --}} In recent years, numerous
studies of the nonequilibrium dynamics of carriers in superconductors have
been performed using femtosecond time-resolved spectroscopy
\cite{kabanov_raten, carr_pp, kaindl_science, kaindl_prb, gedik, gedik2,
  kabanov_lsco}. In a typical experiment, the sample is excited with an
intense fs laser pulse (pump pulse), and after a delay time $\Delta t$,
spectra are measured using a second, less intense, laser pulse (probe pulse).
As the nature of the interactions between quasiparticles in the high-$T_c$
cuprates is still under debate \cite{norman_review}, it is interesting to
directly observe the characteristic dynamics of condensate depletion and
Cooper-pair recombination. This can be done with real-time optical techniques.
In the high-$T_c$ superconductor Bi$_2$Sr$_2$CaCu$_2$O$_{8+\delta}$ (BSCCO),
for example, relaxation times of about 50 ps have been measured
\cite{kaindl_prb}; the observed decay is two-component (biexponential).

Theoretical attempts to model these experiments have, on the one hand, used
quasi-equilibrium models (so-called $\mu^*,\,T^*$ models,
\cite{nicol_carbotte}) to describe the state excited by the pump pulse. On the
other hand, rate equation approaches based on the phenomenological
Rothwarf-Taylor model \cite{rtaylor} have been used
\cite{kabanov_raten,kaindl_prb,kabanov_lsco} to describe the recovery dynamics
of the superconducting state. It is assumed that the dynamics are governed by
the creation of high-energy phonons due to Cooper-pair recombination and
subsequent phonon decay. So far, there has been no attempt to describe the
excitation \emph{and} relaxation dynamics on equal footing. As well, no
microscopic description of the related time dynamics is available.

In this Letter, we present a theory which can describe the femtosecond
excitation and relaxation processes from a microscopic viewpoint. In
particular, we consider high-$T_c$ cuprates, using a realistic band structure
and considering coupling to two important phonon modes (breathing and buckling
modes, which are strongly coupled to the superconducting CuO$_2$ planes
\cite{phonons_exp}).  We employ the approach of density-matrix theory, which
has been used to some extent to describe ultrafast dynamics in semiconductors
(see e.g. Ref. \cite{kuhn_dm_theory,kira_koch,marten}).
    
\textit{{\label{sec:theory}Theory --}} We start from a Hamiltonian
$H=H_{\text{sc}}+H_{\text{field}}+H_{\text{phon}}$, where $H_{\text{sc}}$
describes the superconducting state, $H_{\text{field}}$ gives the interaction
with the classical electromagnetic field, and $H_{\text{phon}}$ models the
bare phonons and their interaction with the electrons. Explicitly we write
      \begin{align}
         \label{eq:Hamiltonian_Hsc}
         &H_{\text{sc}}=\sum_{\vec{k}s}(\epsilon_{\vec{k}}\!-\!\mu)c^+_{\vec{k}s}c_{\vec{k}s}
         + \sum_{\vec{k}}\left(\Delta_{\vec{k}}c^+_{\vec{k}\auf}c^+_{-\vec{k}\ab}+\text{h.c.}\right),\\
         \begin{split}
            \label{eq:Hamiltonian_Hfield}
           &H_{\text{field}}
           =-\frac{e\hbar}{m}\sum_{\vec{k}\vec{q}s}(\vec{k}\cdot\vec{A}_{\vec{q}})c^+_{\vec{k}+\frac{\vec{q}}{2}s}c_{\vec{k}-\frac{\vec{q}}{2}s}\\
           &\quad +\frac{e^2}{2m}\sum_{\vec{k}s}(\vec{A}_{\vec{q}-\vec{k}}\cdot\vec{A}_{\vec{q}})c^+_{\vec{k}s}c_{\vec{k}s},\text{ and}
         \end{split}\\
         \begin{split}
            \label{eq:Hamiltonian_Hphon}
           &H_{\text{phon}}=\sum_{\vec{q}j}\hbar\omega_{\vec{q}j}\left(b^+_{\vec{q}j}b_{\vec{q}j}+\frac{1}{2}\right)\\
           &\quad +\sum_{\vec{p}j\vec{k}s}\left(g_{\vec{p}\vec{k}js}(b^+_{-\vec{p}j}+b_{\vec{p}j})c^+_{\vec{k}+\vec{p},s}c_{\vec{k}s}+\text{c.c.}\right).
         \end{split}
    \end{align}
    In Eq. (\ref{eq:Hamiltonian_Hsc}), $\epsilon_{\vec{k}}$ is a tight-binding
    band structure as measured by Kordyuk \textit{et al.}
    \cite{bandstr_kordyuk}, $\mu$ is the chemical potential, and
    $\Delta_{\vec{k}}=\Delta_0(\cos k_x - \cos k_y)/2$ denotes a $d$-wave
    order parameter with $\Delta_0=30$ meV. $c^+$ and $c$ are the electronic
    creation and annihilation operators, respectively. The $j$ index counts
    the different phonon modes. In Eq. (\ref{eq:Hamiltonian_Hfield}),
    $\vec{A}_{\vec{q}}$ denotes the Fourier component of the vector potential
    the superconductor interacts with. It includes both the pump and probe
    fields; as the interaction with the pump field is nonlinear, the quadratic
    terms in $\vec{A}$ are needed. $H_{\text{phon}}$ in Eq.
    (\ref{eq:Hamiltonian_Hphon}) includes the bare phonons, having the
    dispersion $\omega_{\vec{q}j}$, and the electron-phonon interaction,
    described by the coupling matrix elements $g_{\vec{p}\vec{k}js}$. $b^+$
    and $b$ are the creation and annihilation operators for the phonons. We
    consider the important breathing and buckling phonon modes
    \cite{elph_coupling}.  These two modes are thought to be most strongly
    coupled to the superconducting state, and thus the most relevant for
    scattering processes which can lead to relaxation of exited
    quasiparticles.

    We first perform a Bogoliubov transformation
    $\alpha^+_{\vec{k}}=u_{\vec{k}}c^+_{\vec{k}\auf}-v_{\vec{k}}c_{-\vec{k}\ab}$,
    $\beta^+_{\vec{k}}=u_{\vec{k}}c_{-\vec{k}\ab}+v_{\vec{k}}c^+_{\vec{k}\auf}$.
    Within the Heisenberg picture we calculate equations of motion for the
    Bogoliubov quasiparticle densities
    $\Av{\alpha^+_{\vec{k}_1}\alpha_{\vec{k}_2}}(t),\Av{\beta^+_{\vec{k}_1}
      \beta_{\vec{k}_2}}(t)$, which correspond to the excited states of a
    superconductor, and the anomalous expectation values
    $\Av{\alpha^+_{\vec{k}_1}\beta^+_{\vec{k}_2}}$, which correspond to the
    condensate of Cooper-pairs. The current density is then given by
    \begin{equation}
     \label{eq:current_density}
     \begin{split}
       \raisetag{2cm}
       & \Av{\vec{j}}(\vec{q},t)=\frac{e\hbar}{m}\sum_{\vec{k}}(2\vec{k}-\vec{q})\\
&\quad \times \left[(u_{\vec{k}+\vec{q}}u_{\vec{k}}+v_{\vec{k}+\vec{q}}v_{\vec{k}})\left(\Av{\alpha^+_{\vec{k}+\vec{q}}\alpha_{\vec{k}}}-\Av{\beta^+_{\vec{k}}\beta_{\vec{k}+\vec{q}}}\right)\right.\\
&+\left.(u_{\vec{k}+\vec{q}}v_{\vec{k}}-v_{\vec{k}+\vec{q}}u_{\vec{k}})\left(\Av{\alpha^+_{\vec{k}+\vec{q}}\beta^+_{\vec{k}}}-\Av{\alpha_{\vec{k}}\beta_{\vec{k}+\vec{q}}}\right)\right]\\
&-\frac{e^2}{2m}\sum_{\vec{k}}\vec{A}_{\vec{q}-\vec{k}}\left(2v_{\vec{k}}^2-\frac{\epsilon_{\vec{k}}}{E_{\vec{k}}}\left(\Av{\alpha^+_{\vec{k}}\alpha_{\vec{k}}}+\Av{\beta^+_{\vec{k}}\beta_{\vec{k}}}\right)\right).
    \end{split}
    \end{equation}
    The first and last terms include Bogoliubov quasiparticle densities, thus
    describing the contribution of the normal part in a two-fluid-model. The
    second term, including anomalous expectation values, describes the
    condensate response. Both Bogoliubov quasiparticle densities and anomalous
    expectation values can be calculated for a given delay time $\Delta t$. As
    the probe field $E_{\text{probe}}=-i\omega A_{\text{probe}}$ is known, the
    optical conductivity $\sigma$ can be calculated via
    $\Av{j}(\vec{q},\omega)=-i\omega\sigma(\vec{q},\omega)A_{\text{probe}}
    (\vec{q},\omega).$ Only the $\vec{q}$-independent conductivity
    $\sigma(\vec{q}\rightarrow 0,\omega)$ will be considered.
 \begin{figure}
 \includegraphics[scale=0.5]{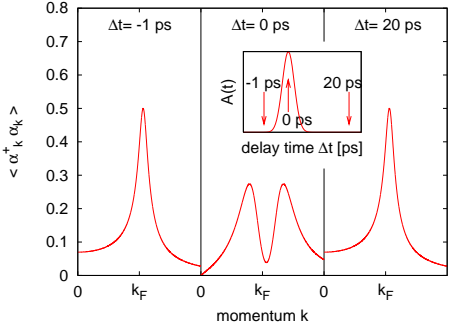}
 \caption{\label{fig:dynamics_kspace}Plot of the excitation process. Starting
   with an equilibrium quasiparticle distribution
   $\Av{\alpha^+_{\vec{k}}\alpha_{\vec{k}}}$ before the pump pulse (left
   panel), a nonequilibrium distribution is excited (middle panel), which then
   relaxes back into equilibrium (right panel). The inset shows the exiting
   pulse, a 50 fs Gaussian. The $\vec{k}$-vectors lie in the 2D CuO$_2$ plane;
   the plot shows a cut along the antinodal direction, from $(0,\pi)$ to
   $(\pi,\pi)$, crossing the Fermi level at $k_F$. The temperature is 4 K.}
 \end{figure}

 \textit{\label{subsec:eqns_of_motion}Equations of motion --} In order to
 calculate $\sigma$, the equations of motion for the Bogoliubov quasiparticle
 distributions and anomalous expectation values have to be solved. They both
 couple to phonon-assisted quantities e.g.
 $\Av{\alpha^+_{\vec{k}_1+\vec{q}}\alpha_{\vec{k}_2}(b^+_{-\vec{q}j}+b_{\vec{q}j})}$.
 We now use second-order cluster expansion
 \cite{cluster_expansion,kuhn_dm_theory} and calculate equations of motion for
 the phonon-assisted quantities, which couple to 4-point quantities such as
 $\Av{\alpha^+_{\vec{k}+\vec{q}}\alpha^+_{\vec{k}}\alpha_{\vec{k}+\vec{q}}
   \alpha_{\vec{k}}}$.  At this point, the hierarchy is broken down by
 factorizing the 4-point quantities. The phonons are assumed to remain
 equilibrated (bath approximation, $\Av{b^+_{\vec{q}j}b_{\vec{q}j}}\rightarrow
 n_{\vec{q}j}$ with the Bose distribution $n_{\vec{q}j}$) while the
 quasiparticles are excited and relax. The equations for the phonon-assisted
 quantities can then be solved, giving rise to a system of
 integro-differential equations. For example, the equation for the Bogoliubov
 quasiparticle occupation $\Av{\alpha^+_{\vec{k}}\alpha_{\vec{k}}}$ reads:
    \begin{equation}
      \label{eq:integro_diff}
      \begin{split}
        \raisetag{2.5cm}
        &\partial_t\Av{\alpha^+_{\vec{k}}\alpha_{\vec{k}}}=-\frac{ie}{m}\vec{k}\cdot\vec{A}_{\vec{q}}M_{\vec{k}\vec{q}}\left(\Av{\alpha_{\vec{k}}\beta_{\vec{k}}}-\Av{\alpha^+_{\vec{k}}\beta^+_{\vec{k}}}\right)\\
        &+\sum_{\vec{qj}}\int_{0}^{\infty}
        ds\frac{\pi|g_{\vec{q}j}|^2}{\hbar^2}\left[(1+n_{\vec{q}j})e^{i(\omega_{\vec{k}+\vec{q}}-\omega_{\vec{k}}+\omega_{\vec{q}j})s}
        \right.\\
        &\quad \times L_{\vec{k}\vec{q}}u_{\vec{k}}u_{\vec{k}+\vec{q}}\Av{\alpha^+_{\vec{k}}\alpha_{\vec{k}}}(t-s)(1-\Av{\alpha^+_{\vec{k}+\vec{q}}\alpha_{\vec{k}+\vec{q}}}(t-s))\\
        &-n_{\vec{q}j}e^{i(\omega_{\vec{k}+\vec{q}}-\omega_{\vec{k}}-\omega_{\vec{q}j})s}
        \\
        & \quad \times L_{\vec{k}\vec{q}}u_{\vec{k}}u_{\vec{k}+\vec{q}}
        \Av{\alpha^+_{\vec{k}+\vec{q}}\alpha_{\vec{k}+\vec{q}}}(t-s)(1-\Av{\alpha^+_{\vec{k}}\alpha_{\vec{k}}}(t-s))
        \\
        &+e^{i(\omega_{\vec{k}+\vec{q}}+\omega_{\vec{k}}-\omega_{\vec{q}j})s}
        \\
        &\quad \times \left.
          M_{\vec{k}\vec{q}}u_{\vec{k}+\vec{q}}v_{\vec{k}}\Av{\beta^+_{\vec{k}+\vec{q}}\beta_{\vec{k}+\vec{q}}}(t-s)\Av{\alpha^+_{\vec{k}}\alpha_{\vec{k}}}(t-s)\right]
      \end{split}
     \end{equation}
     with
     $L_{\vec{k}\vec{q}}=u_{\vec{k}+\vec{q}}u_{\vec{k}}+v_{\vec{k}+\vec{q}}v_{\vec{k}}$,
     $M_{\vec{k}\vec{q}}=u_{\vec{k_+\vec{q}}}v_{\vec{k}}-v_{\vec{k}+\vec{q}}u_{\vec{k}}$
     being the relevant matrix elements, and
     $\omega_{\vec{p}}=E_{\vec{p}}/\hbar$, where
     $E_{\vec{p}}=\sqrt{\epsilon_{\vec{p}}^2+\Delta_{\vec{p}}^2}$ is the
     Bogoliubov quasiparticle dispersion. There are 4 equations for the 4
     expectation values appearing in Eq. (\ref{eq:current_density}), all with
     a similar structure. The full system is published elsewhere
     \cite{not_yet_dm}. On this level, the equations are similar to the ones
     obtained within the Keldysh formalism, with the difference that here the
     \emph{nonequilibrium} distributions
     $\Av{\alpha^+_{\vec{k}}\alpha_{\vec{k}}}(t-s)$ with their full
     time-dependences contribute.
   
     By using the Markovian approximation \cite{markov}, the integrals can be
     solved and one finds, for example,
    \begin{equation}
      \label{eq:Boltzmann_eqn}
      \begin{split}
        &\partial_t\Av{\alpha^+_{\vec{k}}\alpha_{\vec{k}}}=-\frac{ie}{m}\vec{k}\cdot\vec{A}_{\vec{q}}M_{\vec{k}\vec{q}}\left(\Av{\alpha_{\vec{k}}\beta_{\vec{k}}}-\Av{\alpha^+_{\vec{k}}\beta^+_{\vec{k}}}\right)\\
&+\sum_{\vec{qj}}\frac{\pi|g_{\vec{q}j}|^2}{\hbar^2}\left(\Gamma^{(1)}_{\vec{k}\vec{q}j}\Av{\alpha^+_{\vec{k}}\alpha_{\vec{k}}}(1-\Av{\alpha^+_{\vec{k}+\vec{q}}\alpha_{\vec{k}+\vec{q}}})\right.\\
&\quad-\Gamma^{(2)}_{\vec{k}\vec{q}j}\Av{\alpha^+_{\vec{k}+\vec{q}}\alpha_{\vec{k}+\vec{q}}}(1-\Av{\alpha^+_{\vec{k}}\alpha_{\vec{k}}})\\
&\quad \left.-\Gamma^{(3)}_{\vec{k}\vec{q}j}\Av{\beta^+_{\vec{k}+\vec{q}}\beta_{\vec{k}+\vec{q}}}\Av{\alpha^+_{\vec{k}}\alpha_{\vec{k}}}\right),
\end{split}
\end{equation}
with\\
$\Gamma^{(1)}_{\vec{k}\vec{q}j}=(1+n_{\vec{q}j})u_{\vec{k}+\vec{q}}u_{\vec{k}}L_{\vec{k}\vec{q}}\delta(\omega_{\vec{k}+\vec{q}}-\omega_{\vec{k}}+\omega_{\vec{q}j})$,\\
$\Gamma^{(2)}_{\vec{k}\vec{q}j}=n_{\vec{q}j}u_{\vec{k}+\vec{q}}u_{\vec{k}}L_{\vec{k}\vec{q}}\delta(\omega_{\vec{k}+\vec{q}}-\omega_{\vec{k}}-\omega_{\vec{q}j})$,\\
$\Gamma^{(3)}_{\vec{k}\vec{q}j}=u_{\vec{k}+\vec{q}}v_{\vec{k}}M_{\vec{k}\vec{q}}\delta(\omega_{\vec{k}+\vec{q}}-\omega_{\vec{k}}+\omega_{\vec{q}j}).$
This is a Boltzmann-type equation describing both in- and out-scattering with
phonons ($\Gamma^{(1)}_{\vec{k}\vec{q}j},\,\Gamma^{(2)}_{\vec{k}\vec{q}j}$)
and Cooper-pair recombination ($\Gamma^{(3)}_{\vec{k}\vec{q}j}$) processes.
Finally, numerical solution yields the Bogoliubov quasiparticle distributions
and anomalous expectation values, and thus the optical conductivity.

\textit{\label{sec:results}Results --} Exciting the initial Bogoliubov
quasiparticle distribution
$\Av{\alpha^+_{\vec{k}}\alpha_{\vec{k}}}=f(E_{\vec{k}})=(1+\exp{(E_{\vec{k}}/k_B
  T)})^{-1}$ with a fs pump pulse, a nonequilibrium distribution is created,
as shown in Fig. \ref{fig:dynamics_kspace}. The biggest changes are around the
Fermi energy, and the distribution is clearly non-thermal -- quasiparticle
weight is rearranged and consequently the condensate is also in a non-thermal
state. Because of scattering with phonons, this nonequilibrium distribution
can subsequently relax back into an equilibrated one.

\begin{figure}
 \includegraphics[scale=0.55]{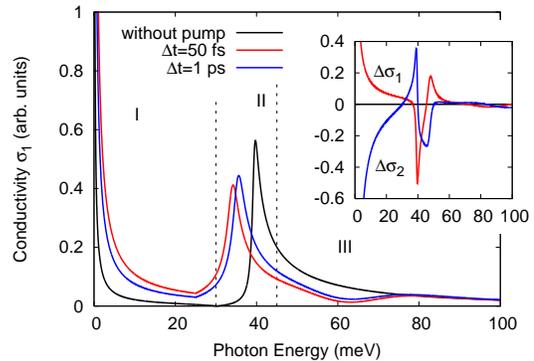}
 \caption{\label{fig:dynamics_espace} Conductivity spectra for buckling (upper
   panel) and breathing (lower panel) modes. The equilibrium spectrum (without
   pump pulse) is shown along with spectra for different delay times. The
   inset shows the change in the real (red) and imaginary (blue) part of the
   conductivity, for $\Delta t=0.5$ ps. It is $\Delta\sigma=\sigma(\Delta
   t)-\sigma_0$ with the equilibrium conductivity $\sigma_0$.}
 \end{figure}

The probe conductivity after the pump and pump-probe spectra are obtained
using the calculated Bogoliubov quasiparticle distributions as shown in Fig.
\ref{fig:dynamics_espace}.
%
%Using the calculated quasiparticle distributions, the probe conductivity after
%the pump and pump-probe spectra can be calculated, as shown in Fig.
%\ref{fig:dynamics_espace}.
%
We can identify three regimes: the low-energy part (I) shows the Drude
response, i.e. the response of the normal part in a two-fluid-model. The
low-frequency power laws for $d$-wave superconductors are still obeyed after
excitation. As Cooper-pairs are broken up by the pump pulse, thus generating
Bogoliubov quasiparticles, the Drude response gets stronger. At higher
energies $\approx 2\Delta_0$ one finds the pair-breaking peak (II). It gets
shifted after pumping, as the superconducting state is depleted and
Cooper-Pairs are broken up. Above the pair-breaking peak (region III), the
absorption, $\alpha\sim\sigma_1/\omega$, is suppressed. In an absorption
process, Cooper-pairs have to be broken up, and the generated quasiparticles
have to have empty states above $2\Delta_0$ to be excited into.  As a large
number of quasiparticles are already excited due to the pump process, there
are less states available than at equilibrium, which decreases the absorption.
So far, only the Drude response part (I) has been measured experimentally
\cite{kaindl_prb}. In principle, however, the regimes II and III could be
measured in THz pump--THz probe experiments.

\begin{figure}
 \includegraphics[scale=0.5]{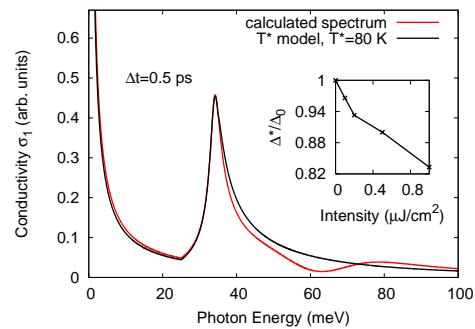}
 \caption{\label{fig:tstar_model} Comparison of a calculated spectrum ($\Delta
   t=0$) with a spectrum calculated using the effective $T^*$ model. $T^*$ is
   chosen in order to fit the calculated position of the pair-breaking peak.
   The inset shows the dependence of the $T^*$ model gap $\Delta^*$ on the
   pump intensity.}
 \end{figure}

The enhancement of the Drude response and the shift of the pair-breaking peak
are also found in a $T^*$ model, where the excited quasiparticle distribution
is assumed to be an equilibrium distribution with an effective temperature
$T^*$ \cite{nicol_carbotte}. However, the suppression above $2\Delta_0$ is
\emph{not} found within a $T^*$ model (see Fig. \ref{fig:tstar_model}). It is
a nonequilibrium effect -- simply speaking, enhancing the temperature does not
create enough Bogoliubov quasiparticles to fill a large number of states above
$2\Delta_0$.

\begin{figure}
 \includegraphics[scale=0.5]{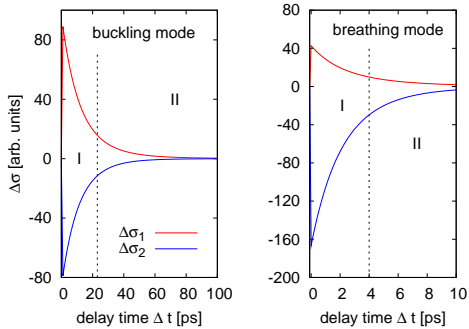}
 \caption{\label{fig:dynamics_tspace} Time evolution of the change in real
   ($\sigma_1$) and imaginary ($\sigma_2$) part of the optical conductivity
   for the buckling (left) and breathing (right) phonon modes. After an
   initial rise due to the pump process, a two-component decay of the
   conductivity follows. In region I scattering and in region II recombination
   dominate, respectively. }
 \end{figure}

Apart from pump-probe spectra, one can also look at the time evolution of the
optical conductivity, which yields additional information about the recovery
dynamics of the superconducting state. Fig. \ref{fig:dynamics_tspace} shows
the change in the conductivity $\Delta\sigma=\sigma(\Delta t)-\sigma_0$, where
$\sigma_0$ is the equilibrium conductivity (without a pump pulse).  $\Delta
\sigma$ initially rises rapidly, as nonequilibrium quasiparticles are created
and the superconducting state is depleted. After pumping, it decays. The
overall timescale of this decay is given by the electron-phonon coupling, and
thus faster for the breathing mode which is more strongly coupled to the
electronic states. The decay is biexponential, with the two timescales
corresponding to quasiparticle-phonon scattering and Cooper-pair
recombination.

 \textit{\label{sec:comparison_rates}Derivation of rate equation approaches
   --} Our microscopic approach can be used to derive a system of rate
 equations, which has been introduced by Kabanov \textit{et al.}
 \cite{kabanov_raten} to describe the combined dynamics of the excited
 quasiparticles and high-frequency phonons. The system is given by
    \begin{equation}
      \label{eq:rate_equations}
      \begin{split}
        &\dot{n}=I_0+\eta N - Rn^2 \\
        &\dot{N}=J_0-\eta\frac{N}{2}+R\frac{n^2}{2}-\gamma(N-N_{T}).
      \end{split}
    \end{equation}
    $n$, $N$ are the numbers of exited quasiparticles and phonons,
    respectively, $\eta$, $R$ are rates denoting pair-breaking and Cooper-pair
    recombination, and $I_0$, $J_0$ are the initial changes in $n$ and $N$.
    $N_{T}$ is the equilibrium phonon number, and $\gamma$ describes phonon
    decay.
    
    So far, we have only considered phonons within the bath approximation,
    where they remain in equilibrium. Our approach can be easily generalized
    to include the nonequilibrium phonon distributions within the Markovian
    approximation. The only modification in Eq.  (\ref{eq:Boltzmann_eqn}) is
    in fact that the phonon distributions $n_{\vec{q}j}$ are then
    time-dependent. A Boltzmann-like equation can also be derived for them.
    With $n\equiv\sum_{\vec{k}}\left(\Av{\alpha^+_{\vec{k}}\alpha_{\vec{k}}}+
      \Av{\beta^+_{\vec{k}}\beta_{\vec{k}}}\right)=2
    \sum_{\vec{k}}\Av{\alpha^+_{\vec{k}}\alpha_{\vec{k}}}$,we can derive an
    equation for $n$ by summing Eq. (\ref{eq:Boltzmann_eqn}) over all
    $\vec{k}$. As only phonon absorption processes, i.e. pair-breaking by
    phonons, are relevant for Eq. (\ref{eq:rate_equations}), only the first
    (initial values) and the last two terms in (\ref{eq:Boltzmann_eqn}) need
    to be considered. The first term gives an initial rate
    $I_0\equiv-\frac{ie}{m}\sum_{\vec{k}}\left[\vec{k}\cdot
      \vec{A}_{\vec{q}}M_{\vec{k}\vec{q}}\left(\Av{\alpha_{\vec{k}}
          \beta_{\vec{k}}}-\Av{\alpha^+_{\vec{k}}\beta^+_{\vec{k}}}\right)\right]$.
    Assuming constant recombination and phonon absorption rates,
    $\Gamma^{(3)}_{\vec{k}\vec{q}j}\rightarrow\Gamma^{(3)}$ and
    $\Gamma^{(2)}_{\vec{k}\vec{q}j}\equiv
    n_{\vec{q}j}\tilde{\Gamma}^{(2)}_{\vec{k}\vec{q}j}\rightarrow
    n_{\vec{q}}\tilde{\Gamma}^{(2)}$, one directly gets the form of Eq.
    (\ref{eq:rate_equations}). A similar calculation with the phonon
    distribution equation yields the second rate equation. Thus, our
    microscopic approach includes the rate equations approach in the limit of
    constant scattering rates $\Gamma^{(i)}$. We can then write the rates $R$
    and $\eta$ as:
     \begin{equation}
       \label{eqn:rates}
         R=\Gamma^{(3)},\,\eta=\tilde{\Gamma}^{(2)}\sum_{\vec{k}}\Av{\alpha^+_{\vec{k}}\alpha_{\vec{k}}}\left(1-\Av{\alpha^+_{\vec{k}}\alpha_{\vec{k}}}\right),
      \end{equation}
     where the rates $\Gamma^{(i)}$ are $\vec{k},\,\vec{q},\,j$-averages of the original
     rates, i.e. $\Gamma^{(i)}=\sum_{\vec{k}\vec{q}j}\Gamma^{(i)}_{\vec{k}\vec{q}j}$.
     
     \textit{\label{sec:summary}Conclusions --} We have utilized
     density-matrix theory to calculate the ultrafast dynamics of high-$T_c$
     superconductors. Our novel microscopic description of the optical
     excitation includes both the depletion of the superconducting condensate,
     as well as relaxation of the excited quasiparticles and Cooper-pair
     recombination due to electron-phonon scattering.
%
%     We have presented a new method for calculating ultrafast dynamics of
%     high-$T_c$ superconductors. Using density-matrix theory, we have
%     developed a microscopic description of the optical excitation, which
%     leads to a depletion of the superconducting condensate, as well as
%     relaxation of the excited quasiparticles and Cooper-pair recombination
%     due to electron-phonon scattering.
%
     Pump-probe spectra, showing nonequilibrium effects above $2\Delta_0$,
     have been calculated as well as the real-time dynamics, where we find a
     biexponential decay produced by quasiparticle-phonon scattering and
     Cooper-pair recombination processes. The relaxation times calculated for
     the buckling modes are compatible with experimental results
     \cite{kaindl_prb}. We have compared our results with spectra calculated
     within the $T^*$ model, finding good agreement in the low-energy limit,
     but our inclusion of nonequilibrium effects yields deviations at higher
     energies. Furthermore, we derive the widely used rate equation approaches
     from our microscopic formalism.  Our method thus provides insight into
     the condensate dynamics of $d$-wave superconductors and includes earlier
     theoretical attempts to describe it.

\textit{\label{sec:acknowledgments}Acknowledgments --}
The authors wish to thank M. Wolf, T. Kampfrath, L. Perfetti, P. Horsch  and
P. Brydon for 
helpful discussions.

%%%%%%%%%%%%%%%%%%%%%%%%%%%%%%%%%%%%%%%%%%%%%%%%%%%%%%%%%%%%%%%%%%%%%%%%%

%%%%%%%%%%%%%%%%%%%%%%%%%%%%%%%%%%%%%%%%%%%%%%%%%%%%%%%%%%%%%%%%%%%%%%%%%%%%

\end{document}